\documentclass[twocolumn,showpacs,superscriptaddress,aps,prl]{revtex4-1}

\usepackage{graphicx}
\usepackage{color}
\usepackage{url}
\usepackage{gensymb}
\usepackage{epsf}

\makeatletter

\@ifundefined{textcolor}{}
{
 \definecolor{BLACK}{gray}{0}
 \definecolor{WHITE}{gray}{1}
 \definecolor{RED}{rgb}{1,0,0}
 \definecolor{GREEN}{rgb}{0,1,0}
 \definecolor{BLUE}{rgb}{0,0,1}
 \definecolor{CYAN}{cmyk}{1,0,0,0}
 \definecolor{MAGENTA}{cmyk}{0,1,0,0}
 \definecolor{YELLOW}{cmyk}{0,0,1,0}
}

\begin{document}

\title{Discovery of unconventional charge density wave at the surface of K$_{0.9}$Mo$_6$O$_{17}$}

\author{Daixiang Mou}
\email{moudaixiang@gmail.com}
\affiliation{Division of Materials Science and Engineering, Ames Laboratory, U.S. DOE, Ames, Iowa 50011, USA}
\affiliation{Department of Physics and Astronomy, Iowa State University, Ames, Iowa 50011, USA}
\author{Aashish Sapkota}
\affiliation{Division of Materials Science and Engineering, Ames Laboratory, U.S. DOE, Ames, Iowa 50011, USA}
\affiliation{Department of Physics and Astronomy, Iowa State University, Ames, Iowa 50011, USA}
\author{H.-H. Kung}
\affiliation{Department of Physics and Astronomy, Rutgers University, Piscataway, NJ 08854, USA.}
\author{Viktor Krapivin}
\affiliation{Department of Physics and Astronomy, Rutgers University, Piscataway, NJ 08854, USA.}
\author{Yun Wu}
\author{A. Kreyssig}
\affiliation{Division of Materials Science and Engineering, Ames Laboratory, U.S. DOE, Ames, Iowa 50011, USA}
\affiliation{Department of Physics and Astronomy, Iowa State University, Ames, Iowa 50011, USA}
\author{Xingjiang Zhou}
\affiliation{National Laboratory for Superconductivity, Beijing National Laboratory for Condensed Matter Physics, Institute of Physics, Chinese Academy of Sciences, Beijing 100190, China}
\author{A. I. Goldman}
\affiliation{Division of Materials Science and Engineering, Ames Laboratory, U.S. DOE, Ames, Iowa 50011, USA}
\affiliation{Department of Physics and Astronomy, Iowa State University, Ames, Iowa 50011, USA}
\author{G. Blumberg}
\affiliation{Department of Physics and Astronomy, Rutgers University, Piscataway, NJ 08854, USA.}
\affiliation{National Institute of Chemical Physics and Biophysics, 12618 Tallinn, Estonia.}
\author{Rebecca Flint}
\author{Adam Kaminski}
\email{kaminski@ameslab.gov}
\affiliation{Division of Materials Science and Engineering, Ames Laboratory, U.S. DOE, Ames, Iowa 50011, USA}
\affiliation{Department of Physics and Astronomy, Iowa State University, Ames, Iowa 50011, USA}

\date{\today}

\begin{abstract}

We use  Angle Resolved Photoemission Spectroscopy (ARPES), Raman spectroscopy, Low Energy Electron Diffraction (LEED)  and x-ray scattering to reveal an unusual electronically mediated charge density wave (CDW) in K$_{0.9}$Mo$_6$O$_{17}$. Not only does K$_{0.9}$Mo$_6$O$_{17}$ lack signatures of electron-phonon coupling, but it also hosts an extraordinary surface CDW, with T$_{S\_CDW}$=220\,K nearly twice that of the bulk CDW, T$_{B\_CDW}$=115\,K. While the bulk CDW has a BCS-like gap of 12\,meV, the surface gap is ten times larger and well in the strong coupling regime. Strong coupling behavior combined with the absence of  signatures of strong electron-phonon coupling indicates that the CDW is likely mediated by electronic interactions enhanced by low dimensionality.
\end{abstract}


\maketitle

Most known CDW materials are mediated by strong electron-phonon (el-ph) interaction \cite{Gruner}, as confirmed by observation of large kinks in the dispersion by ARPES\cite{Valla2000,Valla2004, Yokoya2001, Kiss2007, RossnagelKink}.  
Some of the best known examples are the layered transition-metal dichacogenides and tellurides\cite{Wilson1969,Wilson1975, Moore2010}, where charge order often coexists and competes with superconductivity, due to their common el-ph origin\cite{Wilson1975, Morris1975, Nagata1992, Morosan2006, Wagner2008,Joe2014,Yokoya2001,Kiss2007, Morosan2006}. 
A CDW has been discovered within the pseudogap state of the cuprates\cite{Hoffman2002, Ghiringhelli2012,Chang2012, SilvaNeto2015, Tabis2014, Wu2012, Comin2014, SilvaNeto2014, Hanaguri2004,Vershinin2004}, although its origin remains unclear.  The observation of phonon anomalies suggests el-ph coupling may play a role\cite{LeTacon2014, Reznik2006,Astuto2002}, however, a number of theoretical models suggest that this CDW could be electronically mediated\cite{Kivelson2003, Kivelson2007,Lee2014,Andrey2015}. Electron-electron (el-el) interactions are also thought to be responsible for the CDW found in related cuprate ladder compounds\cite{Blumberg2002}. 

The properties of materials can be dramatically altered at the surface. In CDWs often the transition temperature is enhanced at the surface \cite{Rosen2013}, an effect known as an \emph{extraordinary} transition\cite{McMillan1976, Brown2005}. Recently such effect was also reported in a monolayer\cite{Qing-Yan2012,Liu2012,Xi2015}. The increased $T_C$ has been attributed to enhanced interactions due to the decreased dimensionality\cite{LeeJJ2014,PhysRevB.80.241108}.  In this letter, we show that K$_{0.9}$Mo$_6$O$_{17}$ has an enhanced surface transition temperature, and a surface energy gap an order of magnitude larger than the bulk. Despite the strong coupling nature of the surface order, K$_{0.9}$Mo$_6$O$_{17}$ shows no signatures of strong el-ph coupling, either in the phonon or electronic structure, making it a new candidate for an el-el mediated CDW.

K$_{0.9}$Mo$_6$O$_{17}$ belongs to a family of materials including both one dimensional (1D) and two dimensional (2D) systems \cite{Greenblatt1988} and has been regarded as a model quasi-2D CDW material with T$_{B\_CDW}$ $\sim$ 115\,K \cite{Buder1982}.
Its crystal structure \cite{Vincent1983} consists of a stacking of molybdenum-oxygen slabs (Mo$_6$O$_{17}$) along the hexagonal c axis with potassium atoms intercalated between these slabs.
The Mo-O layers consist of Mo$_2$O$_{10}$ zigzag chains along three directions, and the 2D Fermi surface (FS) can be constructed by superimposing three sets of quasi-1D FS lines, with a weak hybridization.
The quasi-1D character of the FS is likely critical to the unusual properties of this materials as it naturally provides an exceptional nesting condition \cite{WHANGBO1991,Gweon1997}. Indeed, the measured CDW vectors agree well with FS nesting vectors that connect two crossing points of the quasi-1D FS sheets \cite{Escribe-filippini1984,Mallet1999}.

\begin{figure*}
\includegraphics[width=0.8\textwidth]{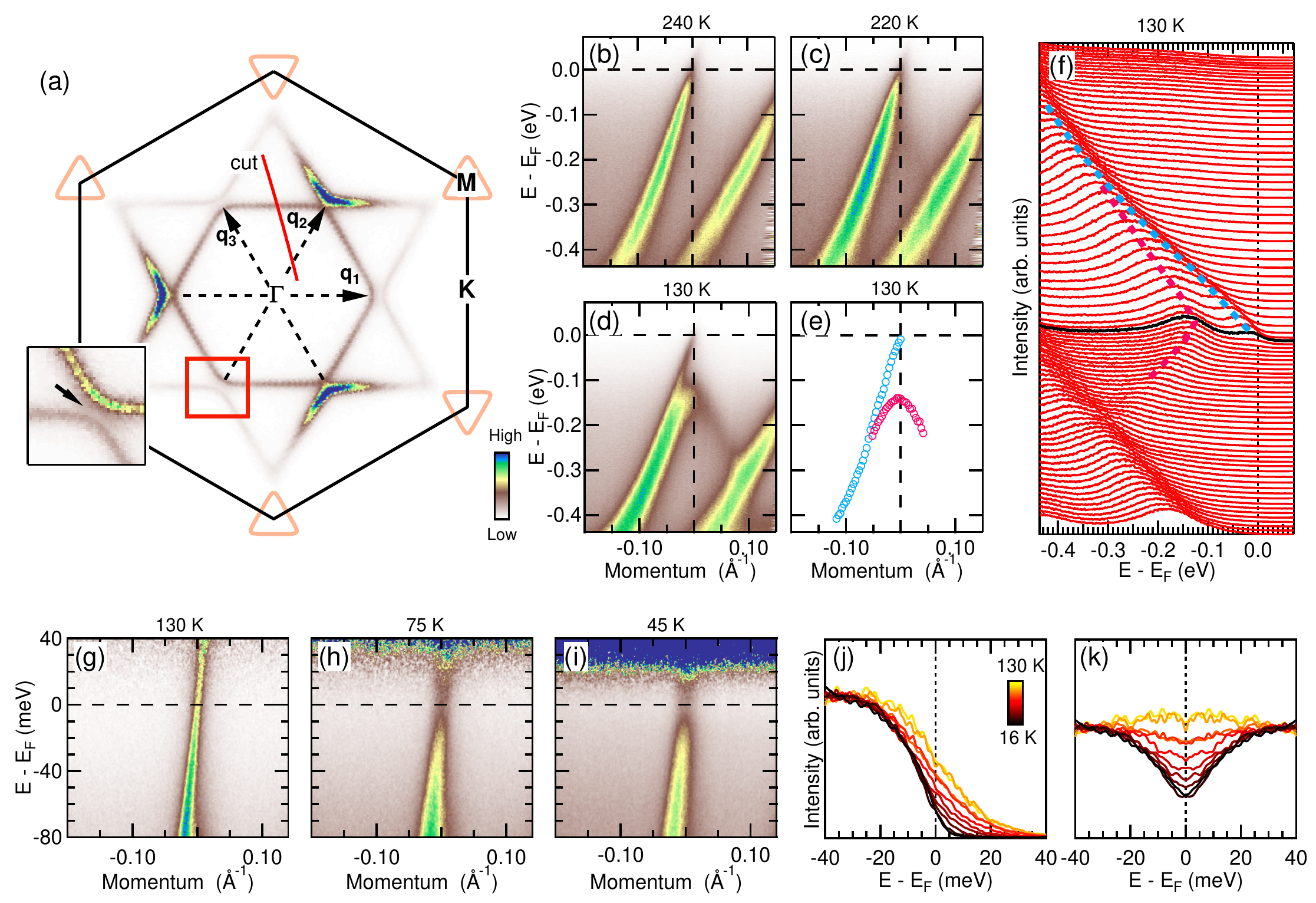}
\caption{Bulk and surface CDW gaps:
(a) Measured FS  at 130\,K. Intensity is integrated in the energy window of  {\it E$_F$} $\pm$ 10\,meV and data are symmetrized with six-fold symmetry. Orange triangles centered at the M point illustrate the hole FS pockets not included in our measurements. Dashed arrows indicate three nesting vectors, each connecting two crossing points of the quasi-1D FS sheets\cite{WHANGBO1991}. The red rectangle is expanded in the left-bottom insert to demonstrate the band hybridization at crossing points of quasi-1D sheets.
(b)-(d) Electronic structure measured along the cut (solid line) shown in (a), at selected temperatures.
(e) Extracted band dispersion of two branches from (d).
(f) Corresponding EDC lines of the electronic structure in (d).
(g)-(i) Detailed electronic structure of the same cut near {\it E$_F$} at 130\,K, 75\,K and 45\,K plotted as ARPES intensity divided by the Fermi function.
(j) Temperature dependence of the EDCs at {\it k$_F$}. All EDCs were normalized using intensities between -40\,meV $\sim$ -30\,meV to show the leading edge shift below T$_{B\_CDW}$.
(k) Same as in (j), but symmetrized about {\it E$_F$}.
}
\label{fig1}
\end{figure*}

In this letter, we show that there is a enhancement of CDW transition temperature and even larger enhancement of the CDW energy gap (by an order of magnitude) on the surface of K$_{0.9}$Mo$_6$O$_{17}$. More surprisingly, we demonstrate that this materials lacks usual signatures of strong el-ph coupling. This combined with large ratio ${2\Delta\over k_BT_C} \sim$ 15 (strong coupling regime) indicates that the CDW is likely mediated by electronic interactions enhanced by low dimensionality.

 K$_{0.9}$Mo$_6$O$_{17}$ single crystals were grown by electrolytic reduction\cite{Xiong2001}.
 The typical size of the samples was $\sim 2\times2\times0.3$\,mm$^3$ in ARPES measurements and $\sim 3\times4\times1$\,mm$^3$ in the x-ray diffraction measurements.
 ARPES measurements were carried out using a laboratory-based system consisting of a Scienta R8000 electron analyzer and a tunable VUV laser light source \cite{Jiang2014}. All data were acquired with a photon energy of 6.7\,eV. The energy resolution of the analyzer was set at 1\,meV and the angular resolution was 0.13$^\circ$ and $\sim$ 0.5$^\circ$ along and perpendicular to the direction of the analyzer slit, respectively. Each temperature dependent data was confirmed by temperature cycling to ensure aging effects did not affect the conclusion.
The high-energy x-ray diffraction experiment was performed using the six-circle diffractometer at the 6-ID-D station at the Advanced Photon Source, Argonne. Synchrotron radiation of 100\,keV with an attenuation length of 3.2\,mm for K$_{0.9}$Mo$_6$O$_{17}$ was used to study the bulk. 
Polarized Raman scattering measurements from the $ab$ surface of the single crystal were performed in quasi-backscattering geometry using the 530.9\,nm excitation line of a Kr$^+$ ion laser with less than 15\,mW of incident power focused to a $50\times100\,\mu m^2$ spot.
The data were corrected for the spectral response of the spectrometer and the CCD. Further technical details are provided in the Supplemental Material.

\begin{figure}
\includegraphics[width=0.8\columnwidth]{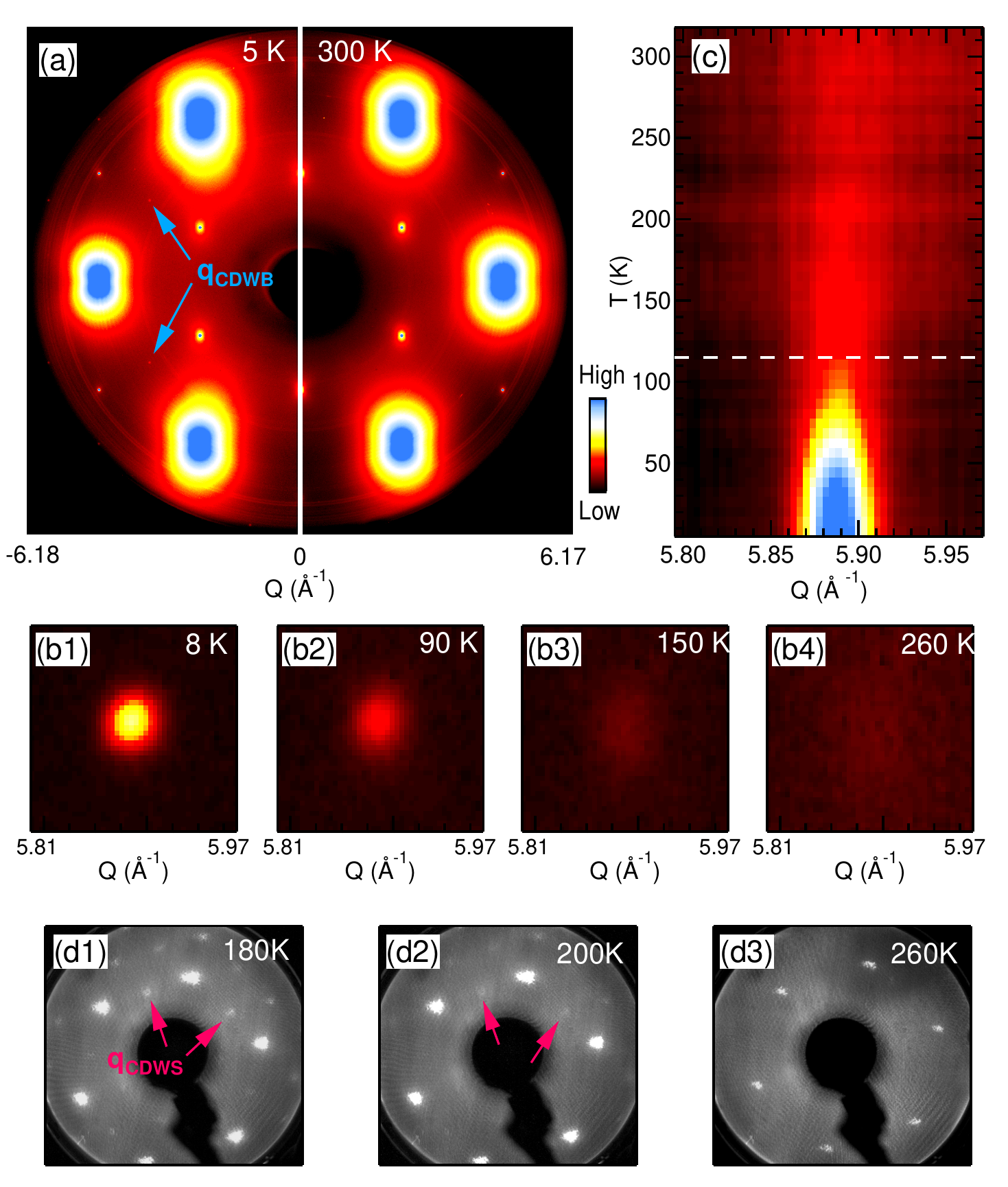}
\caption{Bulk and surface CDW transition.
(a) High-energy x-ray diffraction patterns of the reciprocal lattice plane ({\it H K} 0) obtained at $T$ = 5\,K (left half) and $T$ = 300\,K (right half). The CDW superstructure peaks at low temperature are marked by blue arrows.
(b) High-resolution diffraction patterns of the ($9\over2$ 0 0) CDW peak at selected temperatures.
(c) Contour plot of the temperature dependence of the CDW peak. The intensity is obtained by summing up the high-resolution diffraction patterns of the ($9\over2$ 0 0) peak along the transverse direction in (b), and is plotted along the longitudinal direction. The intensity of CDW peak decreased from $T$ = 6\,K to 101 $\pm$ 6\,K. Further temperature increase yields rapid broadening of the peak observable up to T = 300\,K.
(d) Measured LEED images at selected temperatures. Red arrows indicate the observed CDW superstructure peaks at 180\,K and 200\,K which disappear at 260\,K. }
\label{fig2}
\end{figure}

The band dispersion at temperatures well above the CDW transition is shown along a cut through the inner FS in Fig. 1b and is rather unremarkable. When the sample is cooled down to 220\,K, still above the bulk transition temperature, T$_{B\_CDW}$, an astonishing transformation of the band dispersion occurs (Fig. 1c). The single conduction band present at high temperatures appears to split into two branches, as is even more clear at 130\,K (panel d). One branch follows the high-temperature dispersion, while the other reaches only -150\,meV, then bends back towards higher binding energies marking the presence of an energy gap with its minimum located at the metallic {\it k$_F$} value. The dispersion extracted from low temperature energy distribution curve (EDC) is shown in Fig. 1e.
The appearance of the lower branch coincides with the decrease of the intensity of the other branch that crosses {\it E$_F$}. A detailed analysis of the intensities of each branch is presented in the Supplemental Material Fig. S1. The most natural explanation of this unusual behavior is that the measured band dispersion is a combination of surface and bulk contributions. The electronic structure measured at high temperature, quite surprisingly, must be very similar for both, thus we observe a single band. At lower temperatures, we attribute the conducting branch of the band to the metallic bulk of the crystal and the gapped branch to the surface of the crystal, where the gap is due to a CDW with a transition temperature of 230\,K enhanced from the 115\,K bulk value. Surprisingly, the energy value of the gap minimum of the surface CDW is temperature independent. Instead, the intensity of the gapped surface band increases with decreasing temperature. Such unusual behavior is likely a result of strong coupling, and is similar to cuprates \cite{FedorovWeight}.
\begin{figure}
\includegraphics[width=0.8\columnwidth]{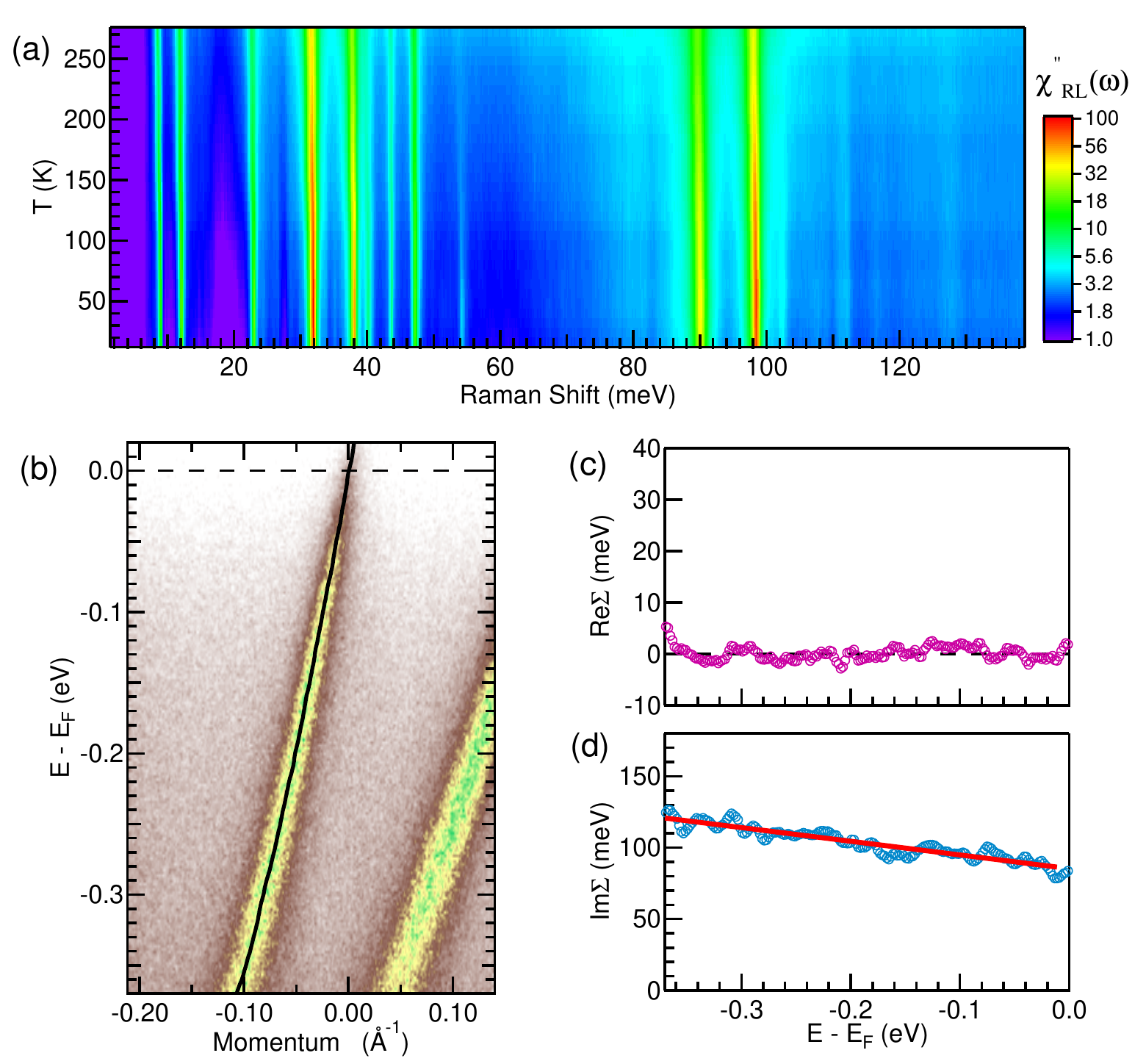}
\caption{Search for signatures of el-ph interaction.
(a) Temperature dependence of the Raman response, $\chi_{Eg}^{\prime\prime}(\omega,T)$, probing the {\it $E_g$} ($D_{3i}$ group) symmetry electronic and phononic excitations. The strong, sharp vertical lines are the 10 bulk phonons of $E_g$ symmetry. The response coding in false color image is in logarithmic scale. The phonon widths at low temperature are close to resolution limited, suggesting high crystal quality.
(b) Band crossing {\it E$_F$} at T = 260\,K. The cut is the same as that in Fig. 1. The parabolic fit to the dispersion is plotted in black.
 (c) Real part of self-energy extracted by subtracting the data and parabolic fit.
 (d)  Imaginary part of self-energy obtained by fitting MDC data with lorentzians, and dividing their half-widths by the velocity. Red solid line is a linear fit.
}
\label{fig3}
\end{figure}

Above the bulk CDW transition, the metallic branch crosses {\it E$_F$} as shown in Fig. 1g-k. Upon cooling below the bulk CDW T$_{B\_CDW}$, the intensity at {\it E$_F$} decreases -  a clear signature of the opening of an energy gap due to the bulk CDW confirmed by raw and symmetrized EDC's (Fig. 1j,k). The temperature at which the bulk gap opens - T$_{B\_CDW}$=115\,K and an energy of $\sim$12 meV are consistent with values expected for the bulk CDW in this material. The leading edge of the EDC's moves to higher binding energies upon cooling, in contrast with the behavior of the large gap at the surface, where the gap energy is temperature independent and instead the gap develops by transferring spectral weight.

To validate these conclusions, we performed extensive low-energy electron diffraction (LEED) and x-ray studies. The LEED studies, carried out with electron energies of 54\,eV, are primarily sensitive to the surface electron density, whereas the x-ray measurements, with energies of 100\,keV, probe the bulk of the sample. In Fig. 2a-c we plot our x-ray data. At 5\,K, we observed additional Bragg peaks, with positions consistent with the previously reported CDW superstructure \cite{Buder1982}. A detailed temperature--dependent measurement of the diffraction peak with high resolution shows that the peaks sharpen and become more intense below the bulk CDW transition temperature T$_{B\_CDW}$ (Fig. 2b and 2c). In the LEED data (Fig. 2d), clear CDW superstructure peaks occur already at 200\,K, much higher than the bulk transition. A significant enhancement of the surface CDW is therefore the most likely explanation of the ARPES data.

To investigate the role of phonons in the formation of the CDW we conducted temperature--dependent measurements of the phonon spectrum using Raman spectroscopy and show the results in Fig. 3a (more detailed plots can be found in the Supplementary Materials Fig. S2). The drop of the electronic background intensity below $\sim$24\,meV and 115\,K indicates the opening of the energy gap, consistent with the ``bulk" ARPES data. However, in contrast to materials where new phonons appear in the CDW phase \cite{Holy1976, Klein1982}, no changes in the phonon energies are observed across both bulk and surface transitions for purple bronze. This absence indicates that any changes in the ionic positions across the CDW transition are likely very small and well below our detection limit. %

Of course, Raman spectroscopy is only sensitive to phonons at the center of the Brillouin zone, and not all phonon modes are Raman-active. To verify our hypothesis that the el-ph coupling is weak and does not play a leading role in formation of CDW in purple bronze, we conducted a detailed study of the ARPES dispersion. This method is very sensitive to the coupling of electrons to all phonon modes, as any significant coupling is visible as kinks in the dispersion. In Fig. 3b, the black line is a parabolic fit to ARPES data over a large energy range and reflects a ``bare", non-interacting dispersion. Surprisingly, there is no evidence of deviations of the data from this line (i e. kinks), signifying the absence of strong el-ph coupling. We use the dispersion data and a parabolic fit to extract the real part of the self energy and the MDC width to obtain the imaginary part, shown in Fig. 3c and 3d respectively. Again, there is no evidence of the coupling of electrons to phonons or any other collective excitations in this material. This is in stark contrast to  conventional CDW materials like NbSe$_2$, where several such features were reported \cite{RossnagelKink}. Based on our ARPES data, any peaks in the real part of the self energy must be smaller than the 3\,meV error bars; by constrast, MgB$_2$ has an $\sim$80\,meV peak \cite{Mou2015} and NbSe$_2$ an $\sim$30\,meV peak\cite{RossnagelKink}. The absence of such features is highly unusual and implies that the el-ph coupling does not play significant role in the formation of the CDW, as suggested by our Raman and ARPES data.


As the combination of ARPES and Raman data seems to rule out el-ph coupling as the origin of the CDW in purple bronze, we must consider electronic mechanisms.  Indeed, el-el interactions drive a CDW in the Sr$_{14}$Cu$_{24}$O$_{41}$ ladder compounds \cite{Blumberg2002} and possibly in the cuprates\cite{Hoffman2002,Kivelson2003, Kivelson2007, Ghiringhelli2012, Lee2014, Andrey2015}.  However, these systems are so strongly interacting as to be magnetic, and as a consequence, the CDW wave-vector is unrelated to nesting.  However, K$_{0.9}$Mo$_6$O$_{17}$ shows no signs of magnetism, and the CDW wave-vector is clearly determined by nesting, making a magnetic origin unlikely.  Indeed, the FS of K$_{0.9}$Mo$_6$O$_{17}$ consists of quasi-1D lines, leading to extremely good nesting, similar to the rare-earth tellurides, which also show a 2D CDW transition\cite{Moore2010}.  The tellurides have strong el-ph coupling\cite{Lavagnini2010,Maschek2015}, but we require an alternative interaction for K$_{0.9}$Mo$_6$O$_{17}$.  The on-site el-el interaction is repulsive in the CDW channel, however further neighbor interactions are attractive.  Normally one would not expect these in a good metal, however, the quasi-1D nature of the bands reduces the screening of the Coulomb interaction.  Therefore, further neighbor interactions could stabilize a CDW at wave-vectors connecting two of the quasi-1D FSs, consistent with the wave-vectors here. The relevance of quasi-1D physics \cite{Lee1973} is also seen in the power law behavior of $\mathrm{Im} \Sigma (\omega) \sim \omega$. $\mathrm{Im} \Sigma (\omega)$ has been extracted from the data (Fig. 3d) and is linear with energy at least up to 0.4\,eV (highest binding energy measured). The possibility of such interactions stabilizing a CDW was examined in the related quasi-1D Li$_{0.9}$Mo$_6$O$_{17}$, which similarly shows Luttinger liquid behavior \cite{Gweon2004,Chudzinski2012}. Although no CDW forms, the estimated Coulomb parameters put it close to the regime where el-el interactions could induce a CDW. Thus, the CDW in K$_{0.9}$Mo$_6$O$_{17}$ is likely due to el-el interactions enhanced both by strong nesting and quasi-one-dimensionality.  The screening is further reduced at the surface, explaining the surface strong coupling behavior.

\begin{figure}
\includegraphics[width=0.8\columnwidth]{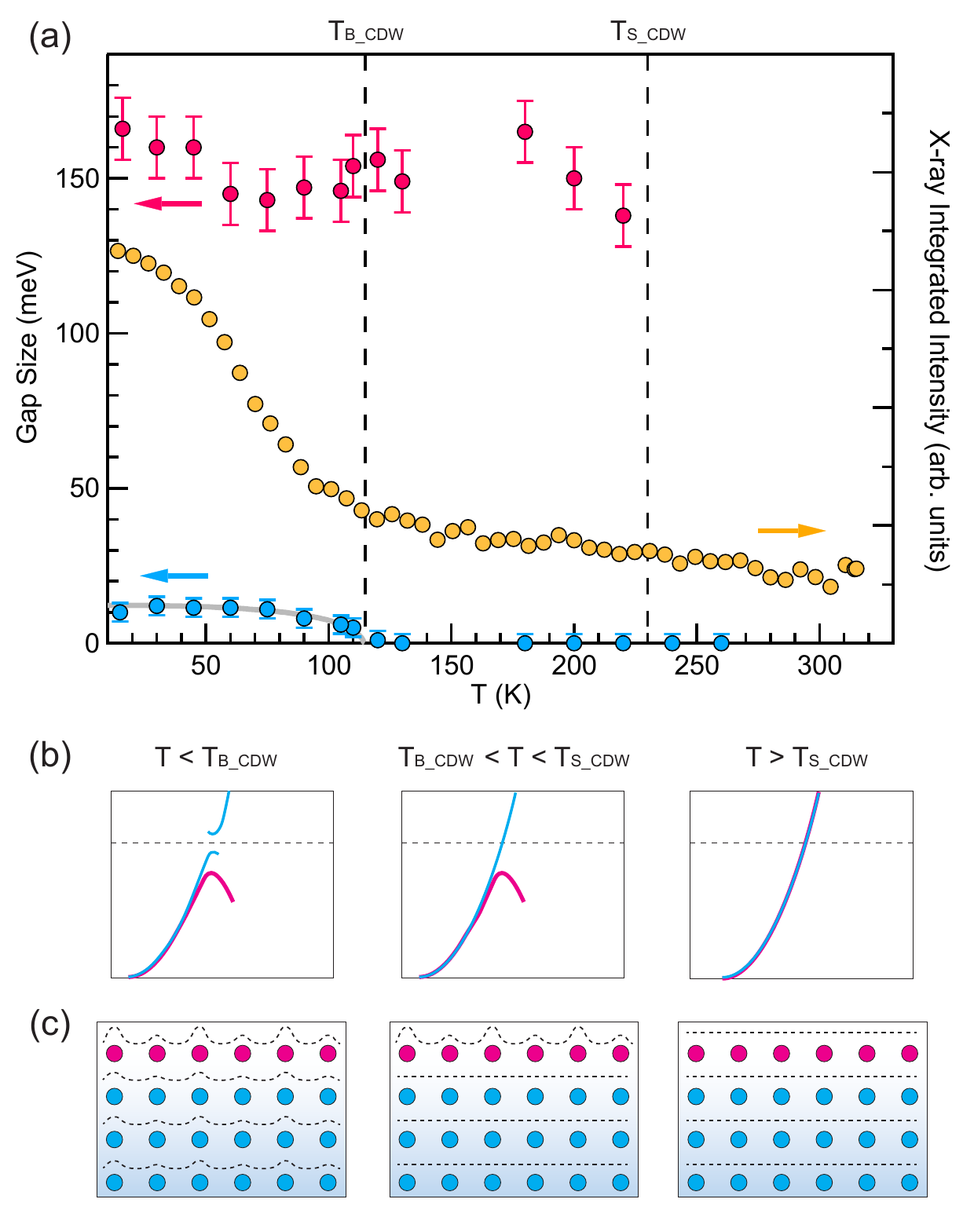}
\caption{Summary of the temperature--dependent CDW gap and band structure evolvement.
 (a) Extracted temperature dependent surface (red solid circles) and bulk (blue solid circles) CDW gap. The surface gap is extracted from the back bending point of the surface band and the bulk gap is extracted from the leading edge shift of {\it k$_F$} EDCs (Fig. 1j). The gray solid line is a BCS-like temperature dependence with $\Delta_0$ = 12\,meV. The integrated intensity of the CDW peak measured by x-ray diffraction (Fig. 2b) is shown with yellow solid circles. The sudden increase of intensity below {\it T$_ {B\_CDW}$} indicates the bulk CDW transition.
 (b) Illustration of the surface (blue line) and bulk (red line) band dispersion in three temperature ranges.
 (c) Illustration of surface (red) and bulk (blue) CDW formation in real space. Dashed lines represent a density distribution of conducting electrons.}
\label{fig4}
\end{figure}

The behavior of the two energy gaps and the bulk CDW order measured by x-ray is shown in Fig. 4a. The bulk CDW gap decreases in BCS-like fashion with temperature and closes at the bulk T$_{B\_CDW}$=115\,K, which corresponds to the dramatic decrease of the intensity of the bulk x-ray superstructure peak. The large energy gap, which we attribute to the CDW at the surface, remains open up to 230\,K. The magnitude of this gap does not change with temperature, instead the spectral weight appears to be transferred to the metallic band that crosses {\it E$_F$}. The ratio $2\Delta\over k_BT_C$ for the bulk band is $\sim$2.5, while at the surface it is in excess of 15. Combining ARPES, LEED and x-ray diffraction results, we illustrate the formation of the surface and bulk CDW and their corresponding band structures in Fig. 4b,c. As the temperature decreases, the surface layer first forms a CDW below T$_{S\_CDW}$ (230\,K) and a large gap (150\,meV) appears in the surface band. The bulk CDW sets in at 115\,K well below T$_{S\_CDW}$ and a much smaller gap (12\,meV) opens in a BCS-like fashion in the bulk band.  Perhaps the most astonishing aspect of our results is that despite such different behavior at the surface and in the bulk, the electronic structures are essentially identical at high temperature. We observe no band splitting at any temperature. By contrast, the single layer material with enhanced T$_{CDW}$ has an electronic structure significantly different from its bulk counterpart \cite{Liu2012,Ugeda2015}. 

In summary we report the discovery of an extraordinary CDW at the surface of purple bronze that lacks any signatures of el-ph coupling and has an energy gap enhanced by more than order of magnitude from the bulk. The strong coupling, combined with dominant role of el-el interaction makes the surface charge order in purple bronze a CDW counterpart to unconventional superconductivity. Indeed, a suppression of this CDW order, if possible, may lead to an exotic superconducting state.

We gratefully acknowledge discussion with Steve Kivelson, Patrick Lee, Mike Norman and Mohit Randeria. Work at the Ames Laboratory was supported by the Department of Energy, Basic Energy Sciences, Division of Materials Sciences and Engineering, under Contract No. DE-AC02-07CH11358 (ARPES measurements). This research used resources of the Advanced Photon Source, a U.S. Department of Energy (DOE) Office of Science User Facility operated for the DOE Office of Science by Argonne National Laboratory under Contract No. DE-AC02-06CH11357 (X-ray scattering measurements). Work at Rutgers was supported by the National Science Foundation under Award NSF DMR-1104884 (Raman measurements). GB acknowledges support from the U.S. Department of Energy, Office of Basic Energy Sciences, Division of Materials Sciences and Engineering under Award DE-SC0005463 (interpretation of material behavior and reconciling information from various techniques).

\bibliography{surfaceCDW}
\newpage 

\section{Supplemental Material for Discovery of unconventional charge density wave at the surface of K$_{0.9}$Mo$_6$O$_{17}$}

\section{SAMPLES}

K$_{0.9}$Mo$_6$O$_{17}$ single crystals were grown by electrolytic reduction.
 K$_2$CO$_3$ and MoO$_3$ mixed power with molar ratio of 1:6 was melt at $\sim$580$^\circ$C in an alumina crucible and an electric current of $\sim$40\,mA was applied through the melt by platinum wire electrodes. Single crystals were obtained on the cathode after three hours.
 The typical size of the samples was $\sim 2\times2\times0.3$\,mm$^3$ in ARPES measurements and $\sim 3\times4\times1$\,mm$^3$ in the x-ray diffraction measurements.

\section{ARPES}

Samples were cleaved \emph{in situ} at a base pressure lower than 8 $\times$ 10$^{-11}$\,Torr. They were cooled using a closed cycle He-refrigerator and  the sample temperature was measured using a silicon-diode sensor mounted on the sample holder. The energy corresponding to the chemical potential was determined from the Fermi edge of a polycrystalline Au reference in electrical contact with the sample.
 Samples were cleaved at three temperatures (40\,K, 130\,K and 260\,K) to check the stability of the surface structure. 
 
 \section{XRAY}

 Two-dimensional diffraction patterns of the ({\it H K 0}) reciprocal plane were measured for overview using a MAR345 detector after rocking the sample through two independent angles up to $\pm$1.8$^\circ$ about axes perpendicular to the incident beam. The sample was mounted on the cold finger of an APD He closed--cycle refrigerator. A detailed study of the temperature dependence of the ($9\over2$ 0 0) CDW Bragg peak was performed using the high-resolution Pixirad detector by rocking the sample through one angle by  $\pm$5$^\circ$. The integrated intensity of the CDW peaks was obtained from fitting two-dimensional Gaussian peaks.

\section{RAMAN}

The sample was cooled with a continuous flow liquid helium cryostat, and the temperatures quoted have been corrected for the laser heating. We used a custom triple-grating spectrometer and liquid nitrogen cooled CCD for analysis and collection of the scattered light. The K$_{0.9}$Mo$_6$O$_{17}$ compound crystallizes in a trigonal structure belonging to the $D_{3i}$ point group (space group P$\overline{3}$, No. 147) \cite{Vincent1983}.
The structure allows a total of 48 phonon modes at the Brillouin zone center ($\Gamma$ point), categorized by the irreducible representations of the $D_{3i}$ group:
$A_u+E_u$ acoustic modes, $13 A_u+13 E_u$ infrared active optical phonons, and $10 A_g+10 E_g$ Raman active phonons. The Raman response function, $\chi^{\prime\prime}(\omega,T)$, measures the electronic and phononic excitations with $E_g$ symmetry ($D_{3i}$ group) when the scattered photon are analyzed in the polarization perpendicular to the incident laser polarization, which is aligned with the crystallographic $a$-axis, whereas excitations with $A_g+E_g$ symmetry ($D_{3i}$ group) are probed when the scattered photon are analyzed in the polarization parallel to the incident laser polarization.

\begin{figure*}
\includegraphics[width=6in]{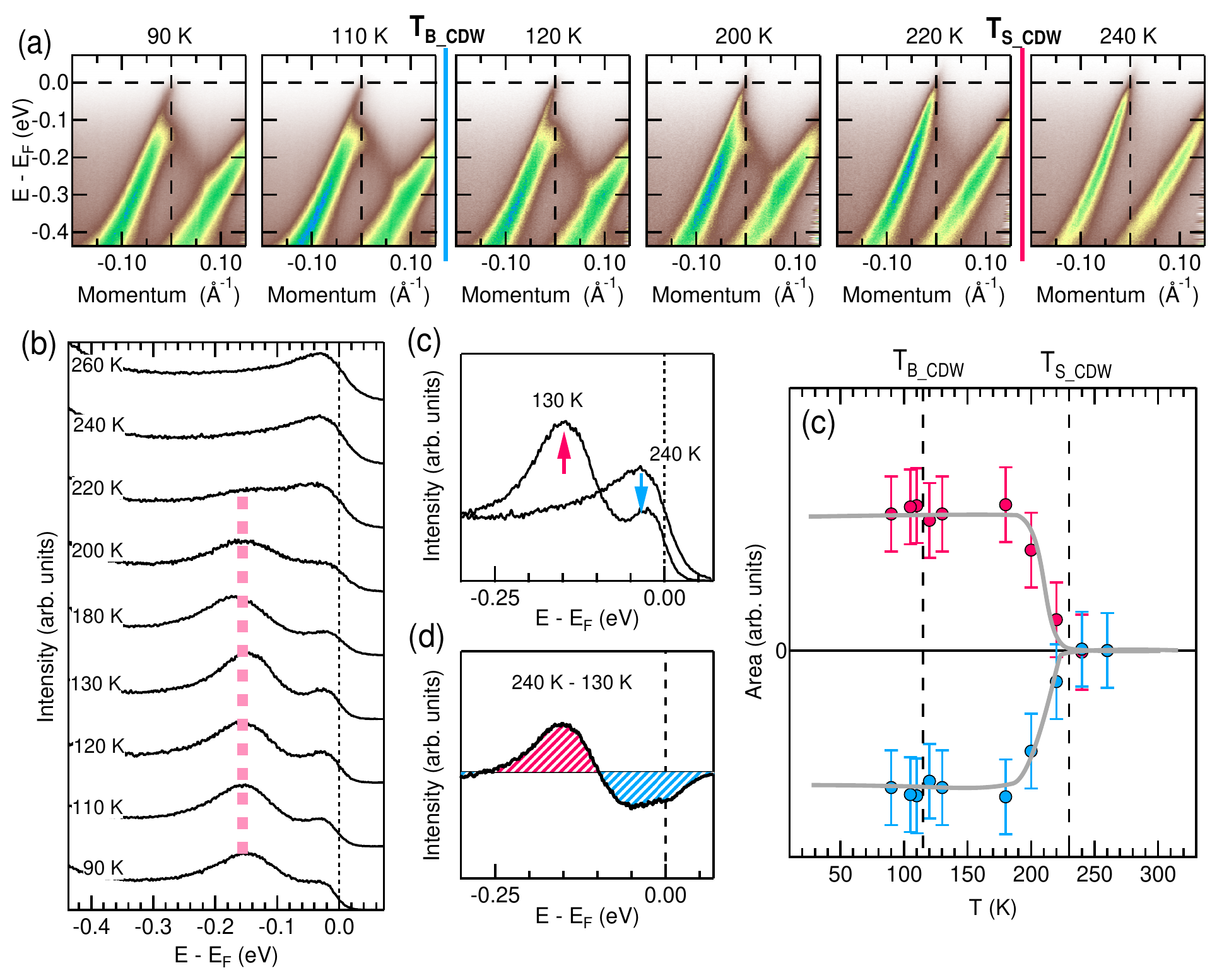}
\caption{Temperature dependent intensity of bulk and surface branches.
(a), Temperature dependence of the electronic structure. Cut position is the same as shown in Fig. 1. {\it T$_{B\_CDW}$}($\sim$115\,K) marks the bulk CDW transition temperature and {\it T$_{S\_CDW}$}($\sim$230\,K) marks the temperature of the onset of the back bending band.
(b), Extracted {\it k$_F$} EDCs at different temperatures. Data are offset vertically for clarity. The peak structure around 150\,meV due to the band back bending disappears at {\it T$_{S\_CDW}$}.
(c), {\it k$_F$} EDC lines at 240\,K and 130\,K. Data are normalized to the total intensity between -0.35\,eV and 0.07\,eV. Two arrows indicate the redistribution of electron density from 240\,K to 130\,K.
(d), Electron density difference between 240\,K and 130\,K of EDCs at {\it k$_F$}.
(e), Electron density redistribution across {\it T$_ {S-CDW}$}.  Data are obtained by adding the positive area around -150\,meV and the negative area around {\it E$_F$} as illustrated in d. The sudden redistribution of electron density from {\it E$_F$} to high binding energy is consistent with the opening of a $\sim$150\,meV gap on the surface band below {\it T$_ {S\_CDW}$}. }
\label{figs1}
\end{figure*}

\begin{figure*}
\includegraphics[width=5in]{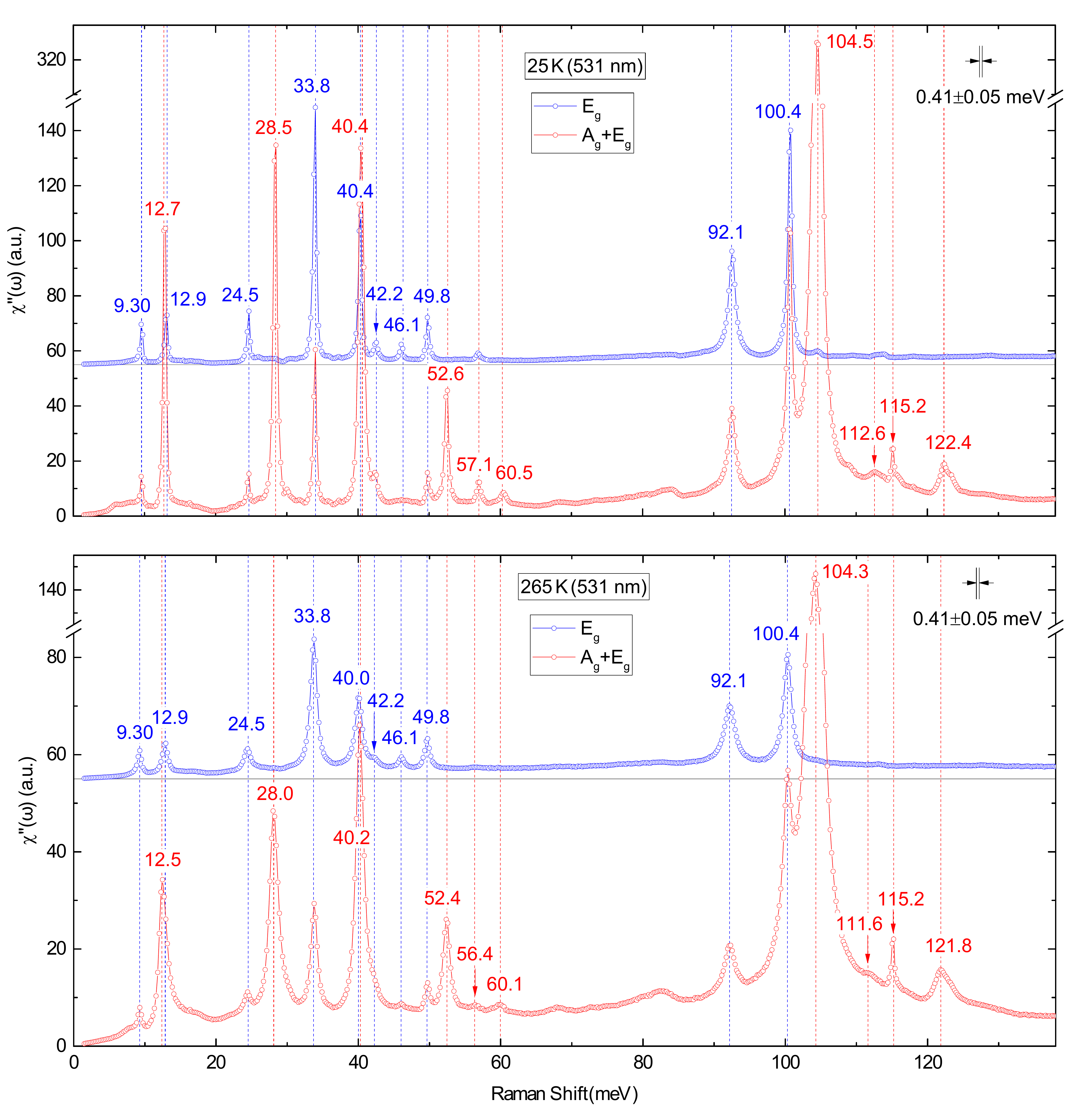}
\caption{Detail analysis of phonon peaks. The Raman response, $\chi^{\prime\prime}(\omega,T)$, probing the $E_g$ (blue) and $A_g+E_g$ (red) symmetries ($D_{3i}$ group) are plot against Raman shift for 265\,K and 25\,K. The data are shifted for clarity. The energies associated with the bulk phonons with $E_g$ and $A_g$ symmetries are labeled next to the phonon peaks in blue and red, respectively.
}
\label{figs2}
\end{figure*}

\end{document}